\documentclass[a4paper,11pt]{article}
\usepackage{pos}

\title{Soft--hard Bayesian inference in heavy-ion collisions}
\ShortTitle{Soft--hard Bayesian inference}
\author*[a,b]{Lipei Du}
\onbehalf{for the JETSCAPE Collaboration}
\affiliation[a]{Department of Physics, University of California,
Berkeley, CA 94720, USA}
\affiliation[b]{Nuclear Science Division, Lawrence Berkeley National Laboratory,
Berkeley, CA 94720, USA}
\emailAdd{ldu2@lbl.gov}

\abstract{
Bayesian analyses of jet quenching commonly constrain jet--medium interaction parameters using a fixed quark-gluon plasma (QGP) background obtained from a soft-sector calibration. This approach neglects uncertainty in the medium and prevents hard observables from providing additional constraints on its properties. We investigate these effects in Pb--Pb collisions at $\sqrt{s_{NN}}=2.76$~TeV using sequential Bayesian inference in a reduced three-parameter space. We vary the initial normalization and nucleon width of the medium together with the jet--medium coupling, combining JETSCAPE medium evolution with MATTER+LBT calculations of high-$p_T$ hadron and inclusive-jet suppression. We first constrain all three parameters using the hard observables alone, then incorporate a projected soft-sector posterior through sequential inference. We find that the hard observables are sensitive to the medium properties and reveal substantial correlations with the jet--medium coupling, reflecting their compensating effects on jet quenching. Propagating the soft-sector information shifts the preferred coupling relative to a fixed-medium extraction, with the update driven primarily by hadron-suppression measurements. Despite substantial differences in the medium parameters and coupling favored by hard-observable-only and sequential inference, their suppression predictions remain similar over much of the measured range. These findings highlight the difficulty of distinguishing changes in the QGP medium from changes in microscopic jet--medium interaction strength using inclusive suppression measurements alone.
}

\FullConference{13th International Conference on Hard and Electromagnetic Probes of High-Energy Nuclear Collisions (HardProbes2026)\\
21-26 June 2026\\
Nashville, Tennessee, USA\\}

\begin{document}
\maketitle

\section{Motivation and Bayesian setup}

Soft observables constrain the QGP medium, whereas hard probes depend on both its evolution and the interactions of energetic partons with it~\cite{jetCollPRC}. Bayesian jet-quenching analyses often use a fixed medium background from an independent calibration, without propagating its uncertainty~\cite{jetBayesHadPRC,globalQhatPRC,hardPRC}. We examine the consequences in Pb--Pb collisions at $\sqrt{s_{NN}}=2.76$~TeV, combining an existing JETSCAPE soft-sector posterior~\cite{softPRC,softPRL} with MATTER+LBT calculations of hadron and jet suppression~\cite{jetMultiPRC,hardPRC}.

We work in a reduced three-parameter space: the TRENTo normalization $N$, controlling the initial density scale; nucleon width $w$, affecting the transverse geometry; and jet--medium coupling $\alpha_s$, which controls the soft-scale interaction strength entering the transport coefficient $\hat q$. Their prior ranges are $N\in[10,20]$, $w\in[0.5,1.5]$~fm, and $\alpha_s\in[0.1,0.5]$. The hard dataset contains 17 hadron and inclusive-jet $R_{AA}$ measurements comprising 206 bins.

At each of 40 design points in $(N,w,\alpha_s)$, we generate event-by-event medium backgrounds using the selected $(N,w)$, then sample and propagate hard partons through them with MATTER+LBT using the corresponding $\alpha_s$~\cite{jetFramework,jetPPPRC,jetMultiPRC}. Gaussian-process emulators interpolate the resulting hadron and jet $R_{AA}$ predictions~\cite{jetBayesMetricPRC}. The Gaussian likelihood includes emulator predictive covariance and diagonal experimental variances, but no additional model-discrepancy term.

Assuming conditional independence of the soft and hard datasets ($D_{\rm S}$ and $D_{\rm H}$) given the model parameters, the sequential posterior in this reduced space is~\cite{duSequentialBayes}
\begin{equation}
 p_{\rm seq}^{\rm red}(N,w,\alpha_s\mid D_{\rm S},D_{\rm H})
 \propto {\cal L}_{\rm H}^{\rm red}(D_{\rm H}\mid N,w,\alpha_s)
 p_{\rm S}(N,w\mid D_{\rm S})\,\pi_J(\alpha_s).
 \label{eq:sequential}
\end{equation}
Here $p_{\rm S}$ is the projection of the full soft-sector posterior onto $(N,w)$, and $\pi_J$ is the prior for $\alpha_s$. The hard data thus update the soft-constrained medium distribution while constraining its correlations with $\alpha_s$. To isolate the role of the soft information, we also perform hard-observable-only inference: all three parameters vary under broad priors using the same hard likelihood, but without imposing $p_{\rm S}$. To represent the conventional fixed-medium approach, we also perform inference with $(N,w)$ held at selected values, varying only $\alpha_s$.

Other medium and jet-sector parameters remain fixed in the hard observable calculations. The fixed jet parameters were obtained from an earlier calibration~\cite{hardPRC} using a different medium background~\cite{dukeMedium} and may not remain optimal in the present setup. The resulting inference isolates the retained parameter correlations rather than providing a complete soft--hard calibration.

\begin{figure}[!htb]
\centering
\includegraphics[width=\linewidth]{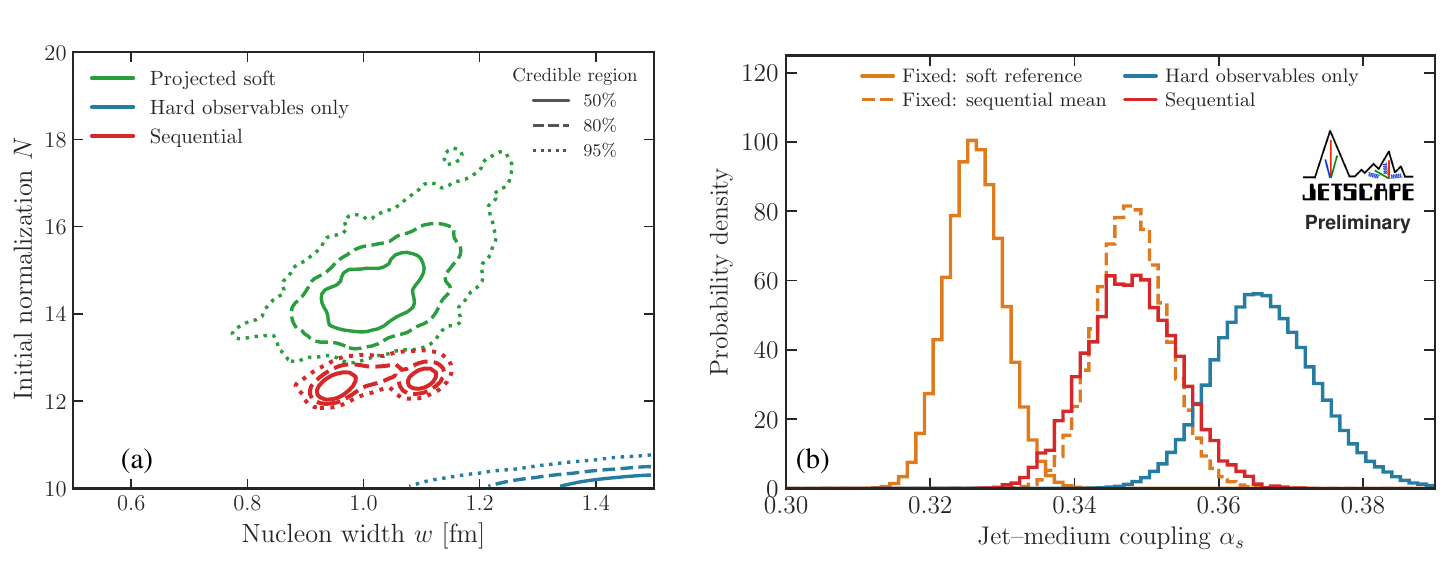}
\caption{(a) Projected soft-sector, broad-prior hard-observable-only, and sequential posteriors in the $(w,N)$ plane. Solid, dashed, and dotted contours enclose 50\%, 80\%, and 95\% credible regions. (b) Coupling posteriors from hard-observable-only and sequential inference, together with fixed-medium extractions at the soft-reference and sequential-mean backgrounds.}
\label{fig:update}
\end{figure}

\section{Results and discussion}

With broad medium priors, hard-observable-only inference favors low $N$ and large $w$ [Fig.~\ref{fig:update}(a)], with means $N=10.17$ and $w=1.39$~fm. Both approach the prior boundaries, indicating preferred directions rather than precise medium determinations. The projected soft posterior instead centers near $(N,w)\simeq(14.7,1.03~\mathrm{fm})$. Combining the constraints, the sequential posterior selects $(N,w)\simeq(12.48,1.01~\mathrm{fm})$ in the low-$N$ tail of the soft distribution. Given the restricted model freedom discussed above, this displacement should not be interpreted as model-independent soft--hard tension.

Fixing the medium at the soft-sector MAP values, $(N,w)=(14.2,1.12~\mathrm{fm})$, gives a posterior mean $\alpha_s=0.326$ with a 16th--84th percentile interval $[0.322,0.330]$. Hard-observable-only and sequential inference yield $0.367$ ($[0.359,0.374]$) and $0.348$ ($[0.342,0.355]$), respectively. 
Fixing the medium at the sequential mean (orange dashed) also gives $\alpha_s\simeq0.348$, but this agreement with the sequential result (red solid) does not imply equivalence between conditioning and marginalization.

\begin{figure}[!ht]
\centering
\includegraphics[width=0.9\linewidth]{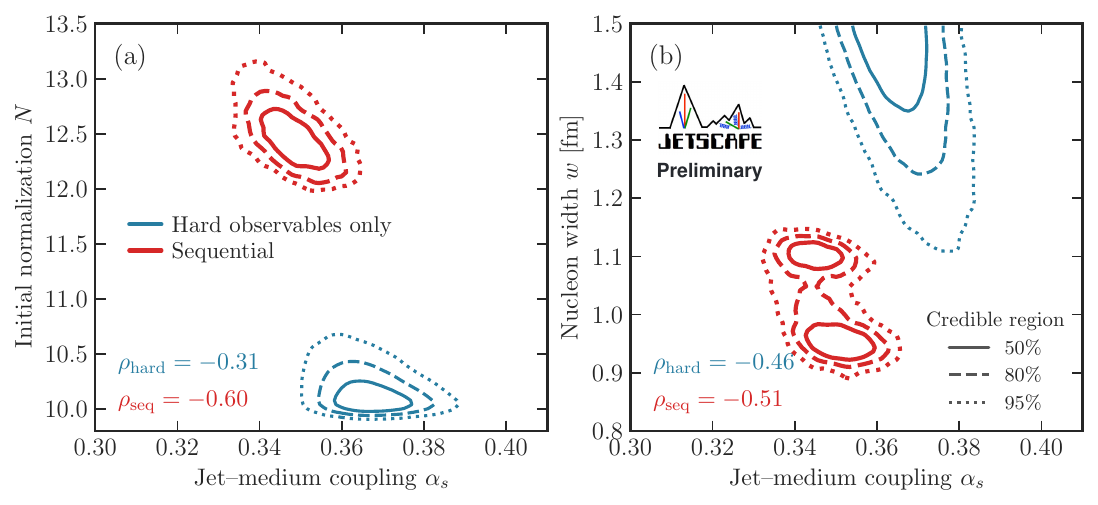}
\caption{Medium--coupling correlations in hard-observable-only and sequential inference: $\alpha_s$ versus (a) $N$ and (b) $w$. Solid, dashed, and dotted contours enclose 50\%, 80\%, and 95\% credible regions; $\rho$ is the Pearson correlation coefficient.}
\label{fig:correlations}
\end{figure}

Figure~\ref{fig:correlations} illustrates the medium--coupling compensation. Increasing $N$ raises the characteristic medium density and temperature, allowing comparable suppression with a smaller $\alpha_s$, and vice versa. This anticorrelation is already present in hard-observable-only inference. In the sequential posterior, $\rho(\alpha_s,N)\simeq-0.60$ and $\rho(\alpha_s,w)\simeq-0.51$. The latter also reflects changes in the transverse geometry and the paths sampled by energetic partons, rather than a simple rescaling of density. Fixing the background removes these compensation directions from the coupling extraction.

Separate sequential updates show that hadron-suppression measurements drive most of the shift toward lower $N$ and the localization of $\alpha_s$, whereas the included jet measurements have a weaker effect on the medium posterior. The steeply falling hadron spectrum makes leading-hadron suppression sensitive to parton energy loss, while reconstructed jets can retain some redistributed shower energy within the jet cone. Different kinematic coverage and experimental uncertainties also contribute, consistent with the observable-selection dependence found in earlier JETSCAPE analyses~\cite{hardPRC,duJetNPA}; this does not establish a universal hierarchy between hadron and jet measurements.

\begin{figure}[!ht]
\centering
\includegraphics[width= 0.9\linewidth]{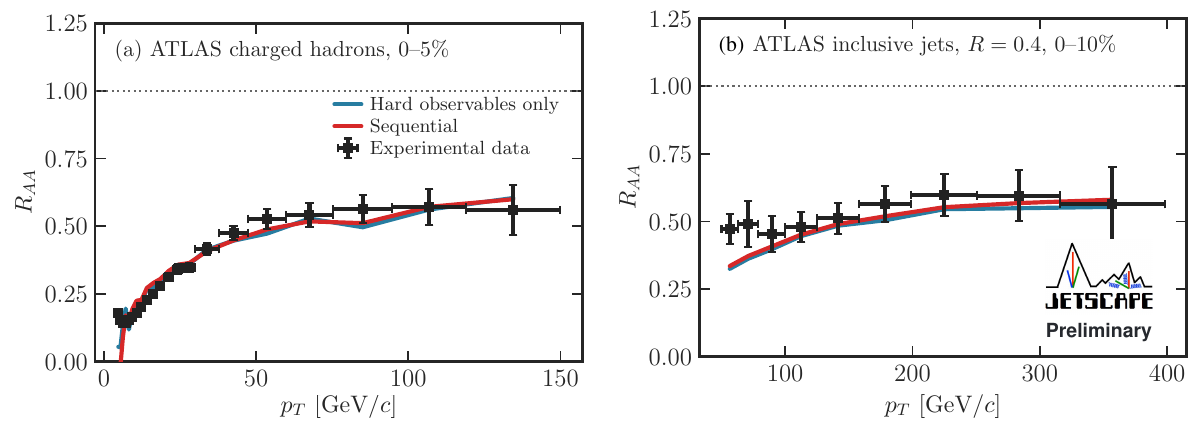}
\caption{Representative hard-observable-only and sequential $R_{AA}$ predictions at $\sqrt{s_{NN}}=2.76$~TeV for (a) ATLAS charged hadrons in 0--5\%~\cite{atlasHadron} and (b) ATLAS inclusive jets with $R=0.4$ in 0--10\%~\cite{atlasJet}. Lines show posterior medians.}
\label{fig:suppression}
\end{figure}

The consequences for measured observables are illustrated in Fig.~\ref{fig:suppression}. Despite substantial changes in $N$, $w$, and $\alpha_s$, hard-observable-only and sequential inference predict similar $R_{AA}$ over much of the displayed kinematic range. The largest visible difference occurs at low $p_T$ for charged hadrons. Inclusive suppression averages over parton production points, trajectories, and shower histories, so distinct medium configurations and interaction strengths can produce comparable net quenching. 
This is the observable manifestation of the correlations in Fig.~\ref{fig:correlations}: inclusive suppression measurements alone cannot easily distinguish changes in the QGP medium from changes in the jet--medium interaction strength.

\section{Conclusions and outlook}

Propagating soft-sector constraints into hard-probe inference changes the preferred medium configurations and jet--medium coupling. Because the hard observables constrain correlated medium--interaction directions, fixing the medium selects one conditional coupling extraction rather than propagating the allowed medium uncertainty. The similar suppression predictions from different posteriors further expose the difficulty of separating medium evolution from microscopic interaction strength. A higher-dimensional calibration will be needed to determine how these findings change when the fixed model parameters are allowed to readjust and to assess the simultaneous description of soft and hard observables.

\section*{Acknowledgments}
This work was supported in part by the U.S. Department of Energy, Office of Science, Office of Nuclear Physics under Contract No. DE-AC02-05CH11231, and by the National Science Foundation through the JETSCAPE Collaboration under Grants No. OAC-2004571 and OAC-2514008. ChatGPT (GPT-6) was used to assist with language polishing.

\end{document}